\documentclass[12pt]{article}
\usepackage{esfconf}

\begin{document}


\title{Holographic Duality, Supersymmetry, and Painlev\'e equation}


\authors{S.V.Ketov\adref{1,2}}


\addresses{\1ad Max-Planck-Institut f\"ur Gravitationsphysik, am M\"uhlenberg 
1, Golm 14476, Germany
\nextaddress
\2ad Institut f\"ur Theoretische Physik, Universit\"at Hannover, Appelstr. 2,
Hannover 30167, Germany}


\maketitle


\begin{abstract}
An example of the holographic correspondence between 2d, N=2 quantum
field theories and classical 4d, N=2 supergravity theories is found. 
The constraints on
 the target space geometry of the 4d, N=2 non-linear sigma-models in N=2 
supergravity background are interpreted as the renormalization flow equations 
in two dimensions. Our geometrical description of the renormalization flow is 
manifestly covariant under reparametrization of the 2d coupling constants. The 
proposed holography is described in terms of the (Weyl) anti-self-dual 
Einstein metrics, whose exact regular (Tod-Hitchin) solutions are governed by 
the Painlev\'e VI equation. 
\end{abstract}



\def\a{\alpha}
\def\b{\beta}
\def\c{\chi}
\def\d{\delta}
\def\e{\epsilon}
\def\f{\phi}
\def\g{\gamma}
\def\h{\eta}
\def\i{\iota}
\def\j{\psi}
\def\k{\kappa}
\def\l{\lambda}
\def\m{\mu}
\def\n{\nu}
\def\o{\omega}
\def\p{\pi}
\def\q{\theta}
\def\r{\rho}
\def\s{\sigma}
\def\t{\tau}
\def\u{\upsilon}
\def\x{\xi}
\def\z{\zeta}
\def\D{\Delta}
\def\F{\Phi}
\def\G{\Gamma}
\def\J{\Psi}
\def\L{\Lambda}
\def\O{\Omega}
\def\P{\Pi}
\def\Q{\Theta}
\def\S{\Sigma}
\def\U{\Upsilon}
\def\X{\Xi}


\def\ve{\varepsilon}
\def\vf{\varphi}
\def\vr{\varrho}
\def\vs{\varsigma}
\def\vq{\vartheta}


\def\ca{{\cal A}}
\def\cb{{\cal B}}
\def\cc{{\cal C}}
\def\cd{{\cal D}}
\def\ce{{\cal E}}
\def\cf{{\cal F}}
\def\cg{{\cal G}}
\def\ch{{\cal H}}
\def\ci{{\cal I}}
\def\cj{{\cal J}}
\def\ck{{\cal K}}
\def\cl{{\cal L}}
\def\cm{{\cal M}}
\def\cn{{\cal N}}
\def\co{{\cal O}}
\def\cp{{\cal P}}
\def\cq{{\cal Q}}
\def\car{{\cal R}}
\def\cs{{\cal S}}
\def\ct{{\cal T}}
\def\cu{{\cal U}}
\def\cv{{\cal V}}
\def\cw{{\cal W}}
\def\cx{{\cal X}}
\def\cy{{\cal Y}}
\def\cz{{\cal Z}}


\def\slpa{\slash{\pa}}                            
\def\slin{\SLLash{\in}}                                   
\def\bo{{\raise-.5ex\hbox{\large$\Box$}}}               
\def\cbo{\Sc [}                                         
\def\pa{\partial}                                       
\def\de{\nabla}                                         
\def\dell{\bigtriangledown}                             
\def\su{\sum}                                           
\def\pr{\prod}                                          
\def\iff{\leftrightarrow}                               
\def\conj{{\hbox{\large *}}}                            
\def\ltap{\raisebox{-.4ex}{\rlap{$\sim$}} \raisebox{.4ex}{$<$}}   
\def\gtap{\raisebox{-.4ex}{\rlap{$\sim$}} \raisebox{.4ex}{$>$}}   
\def\TH{{\raise.2ex\hbox{$\displaystyle \bigodot$}\mskip-4.7mu \llap H \;}}
\def\face{{\raise.2ex\hbox{$\displaystyle \bigodot$}\mskip-2.2mu \llap {$\ddot
        \smile$}}}                                      
\def\dg{\sp\dagger}                                     
\def\ddg{\sp\ddagger}                                   


\def\sp#1{{}^{#1}}                              
\def\sb#1{{}_{#1}}                              
\def\oldsl#1{\rlap/#1}                          
\def\slash#1{\rlap{\hbox{$\mskip 1 mu /$}}#1}      
\def\Slash#1{\rlap{\hbox{$\mskip 3 mu /$}}#1}      
\def\SLash#1{\rlap{\hbox{$\mskip 4.5 mu /$}}#1}    
\def\SLLash#1{\rlap{\hbox{$\mskip 6 mu /$}}#1}      
\def\PMMM#1{\rlap{\hbox{$\mskip 2 mu | $}}#1}   %
\def\PMM#1{\rlap{\hbox{$\mskip 4 mu ~ \mid $}}#1}       %
\def\Tilde#1{\widetilde{#1}}                    
\def\Hat#1{\widehat{#1}}                        
\def\Bar#1{\overline{#1}}                       
\def\bra#1{\left\langle #1\right|}              
\def\ket#1{\left| #1\right\rangle}              
\def\VEV#1{\left\langle #1\right\rangle}        
\def\abs#1{\left| #1\right|}                    
\def\leftrightarrowfill{$\mathsurround=0pt \mathord\leftarrow \mkern-6mu
        \cleaders\hbox{$\mkern-2mu \mathord- \mkern-2mu$}\hfill
        \mkern-6mu \mathord\rightarrow$}
\def\dvec#1{\vbox{\ialign{##\crcr
        \leftrightarrowfill\crcr\noalign{\kern-1pt\nointerlineskip}
        $\hfil\displaystyle{#1}\hfil$\crcr}}}           
\def\dt#1{{\buildrel {\hbox{\LARGE .}} \over {#1}}}     
\def\dtt#1{{\buildrel \bullet \over {#1}}}              
\def\der#1{{\pa \over \pa {#1}}}                
\def\fder#1{{\d \over \d {#1}}}                 


\def\frac#1#2{{\textstyle{#1\over\vphantom2\smash{\raise.20ex
        \hbox{$\scriptstyle{#2}$}}}}}                   
\def\half{\frac12}                                        
\def\sfrac#1#2{{\vphantom1\smash{\lower.5ex\hbox{\small$#1$}}\over
        \vphantom1\smash{\raise.4ex\hbox{\small$#2$}}}} 
\def\bfrac#1#2{{\vphantom1\smash{\lower.5ex\hbox{$#1$}}\over
        \vphantom1\smash{\raise.3ex\hbox{$#2$}}}}       
\def\afrac#1#2{{\vphantom1\smash{\lower.5ex\hbox{$#1$}}\over#2}}    
\def\partder#1#2{{\partial #1\over\partial #2}}   
\def\parvar#1#2{{\d #1\over \d #2}}               
\def\secder#1#2#3{{\partial^2 #1\over\partial #2 \partial #3}}  
\def\on#1#2{\mathop{\null#2}\limits^{#1}}               
\def\bvec#1{\on\leftarrow{#1}}                  
\def\oover#1{\on\circ{#1}}                              

\def\[{\lfloor{\hskip 0.35pt}\!\!\!\lceil}
\def\]{\rfloor{\hskip 0.35pt}\!\!\!\rceil}
\def\Lag{{\cal L}}
\def\du#1#2{_{#1}{}^{#2}}
\def\ud#1#2{^{#1}{}_{#2}}
\def\dud#1#2#3{_{#1}{}^{#2}{}_{#3}}
\def\udu#1#2#3{^{#1}{}_{#2}{}^{#3}}
\def\calD{{\cal D}}
\def\calM{{\cal M}}

\def\szet{{${\scriptstyle \b}$}}
\def\ulA{{\un A}}
\def\ulM{{\underline M}}
\def\cdm{{\Sc D}_{--}}
\def\cdp{{\Sc D}_{++}}
\def\vTheta{\check\Theta}
\def\gg{{\hbox{\sc g}}}
\def\fracm#1#2{\hbox{\large{${\frac{{#1}}{{#2}}}$}}}
\def\ha{{\fracmm12}}
\def\tr{{\rm tr}}
\def\Tr{{\rm Tr}}
\def\itrema{$\ddot{\scriptstyle 1}$}
\def\ula{{\underline a}} \def\ulb{{\underline b}} \def\ulc{{\underline c}}
\def\uld{{\underline d}} \def\ule{{\underline e}} \def\ulf{{\underline f}}
\def\ulg{{\underline g}}
\def\items#1{\\ \item{[#1]}}
\def\ul{\underline}
\def\un{\underline}
\def\fracmm#1#2{{{#1}\over{#2}}}
\def\footnotew#1{\footnote{\hsize=6.5in {#1}}}
\def\low#1{{\raise -3pt\hbox{${\hskip 0.75pt}\!_{#1}$}}}

\def\Dot#1{\buildrel{_{_{\hskip 0.01in}\bullet}}\over{#1}}
\def\dt#1{\Dot{#1}}
\def\DDot#1{\buildrel{_{_{\hskip 0.01in}\bullet\bullet}}\over{#1}}
\def\ddt#1{\DDot{#1}}

\def\Tilde#1{{\widetilde{#1}}\hskip 0.015in}
\def\Hat#1{\widehat{#1}}


\newskip\humongous \humongous=0pt plus 1000pt minus 1000pt
\def\caja{\mathsurround=0pt}
\def\eqalign#1{\,\vcenter{\openup2\jot \caja
        \ialign{\strut \hfil$\displaystyle{##}$&$
        \displaystyle{{}##}$\hfil\crcr#1\crcr}}\,}
\newif\ifdtup
\def\panorama{\global\dtuptrue \openup2\jot \caja
        \everycr{\noalign{\ifdtup \global\dtupfalse
        \vskip-\lineskiplimit \vskip\normallineskiplimit
        \else \penalty\interdisplaylinepenalty \fi}}}
\def\li#1{\panorama \tabskip=\humongous                         
        \halign to\displaywidth{\hfil$\displaystyle{##}$
        \tabskip=0pt&$\displaystyle{{}##}$\hfil
        \tabskip=\humongous&\llap{$##$}\tabskip=0pt
        \crcr#1\crcr}}
\def\eqalignnotwo#1{\panorama \tabskip=\humongous
        \halign to\displaywidth{\hfil$\displaystyle{##}$
        \tabskip=0pt&$\displaystyle{{}##}$
        \tabskip=0pt&$\displaystyle{{}##}$\hfil
        \tabskip=\humongous&\llap{$##$}\tabskip=0pt
        \crcr#1\crcr}}

\def\pl#1#2#3{Phys.~Lett.~{\bf {#1}B} (19{#2}) #3}
\def\np#1#2#3{Nucl.~Phys.~{\bf B{#1}} (19{#2}) #3}
\def\prl#1#2#3{Phys.~Rev.~Lett.~{\bf #1} (19{#2}) #3}
\def\pr#1#2#3{Phys.~Rev.~{\bf D{#1}} (19{#2}) #3}
\def\cqg#1#2#3{Class.~and Quantum Grav.~{\bf {#1}} (19{#2}) #3}
\def\cmp#1#2#3{Commun.~Math.~Phys.~{\bf {#1}} (19{#2}) #3}
\def\jmp#1#2#3{J.~Math.~Phys.~{\bf {#1}} (19{#2}) #3}
\def\ap#1#2#3{Ann.~of Phys.~{\bf {#1}} (19{#2}) #3}
\def\prep#1#2#3{Phys.~Rep.~{\bf {#1}C} (19{#2}) #3}
\def\ptp#1#2#3{Progr.~Theor.~Phys.~{\bf {#1}} (19{#2}) #3}
\def\ijmp#1#2#3{Int.~J.~Mod.~Phys.~{\bf A{#1}} (19{#2}) #3}
\def\mpl#1#2#3{Mod.~Phys.~Lett.~{\bf A{#1}} (19{#2}) #3}
\def\nc#1#2#3{Nuovo Cim.~{\bf {#1}} (19{#2}) #3}
\def\ibid#1#2#3{{\it ibid.}~{\bf {#1}} (19{#2}) #3}


\section{Introduction}

The evidence for the holographic principle (see \cite{ads} for a review) goes 
far beyond the (superconformal) Maldacena conjecture. Certain Quantum Field 
Theories (QFT) apparently allow the dual description in terms of the effective 
gravity theories in higher dimensions. It is, however, not clear {\it where} 
does the holographic correspondence apply and {\it why} does it exist at all?

We give a new (different from the standard AdS/CFT) example of the holographic
 correspondence between 2d, N=2 supersymmetric QFT and classical 4d, N=2 
supergravity theories, by identifying the geometrical constraints on the 
target space geometry of the 4d Non-Linear Sigma-Models (NLSM), in the N=2 
supergravity background, with the RG flow equations in the 2d, N=2 QFT. 
For simplicity, we restrict ourselves to the four-dimensional NLSM target 
spaces, by considering a single NLSM hypermultiplet only. This limited class 
of the 4d, N=2 NLSM describes intergrable deformations of the 2d, N=2 
Superconformal Field Theories (SCFT). The exact RG flow is highly constrained 
by N=2 supersymmetry, being described by the effective regular (Tod-Hitchin)
 metrics governed by the exact solutions to the Painlev\'e VI equation 
\cite{tod,hit}.

The N=2 scalar (hypermultiplet) couplings in the 2d, N=2 supergravity are 
well-known to be described by the NLSM with the quaternionic-K\"ahler target 
spaces of negative scalar curvature \cite{bw}. The four-dimensional 
quaternionic-K\"ahler target space is (Weyl) Anti-Self-Dual (ASD) and Einstein,
 i.e.
$$ W^+_{abcd}=0 \quad{\rm and}\quad  R_{ab}=\ha\L g_{ab}~,\qquad \L=-24\k^2~,
\eqno(1)$$
where $W=W^- + W^+$ is the Weyl tensor, $R_{ab}$ is the Ricci 
tensor, $a,b,c,d=1,2,3,4$, and $\k$ is the gravitational coupling constant.

The R-symmetry of N=2 supersymmetry implies the $SU(2)$ isometry of the NLSM 
metric. The (non-degenerate) action of this isometry in four internal 
dimensions leads to the three-dimensional orbits that can be parametrized by 
the `radial' coordinate $(t)$ to be identified with the RG parameter of the
dual QFT in 2d. The $SU(2)$-invariant metrics are most conveniently described 
in the Bianchi IX formalism of general relaitivity, having manifest $SU(2)$ 
symmetry. Given a `radial' coordinate $r$ and `Euler angles' $(\q,\j,\f)$, the
 $SU(2)$-covariant one-forms are  $\s_1=\ha(\sin\j d\q -\sin\q\cos\j d\f)$, 
$\s_2=-\ha(\cos\j d\q +\sin\q\sin\j d\f)$ and $\s_3=\ha(d\j+\cos\q d\f)$, being
subject to the relation $\s_i\wedge \s_j=\ha \ve_{ijk}d\s_k$. 

The simplest {\it symmetric} quaternionic space, which is relevant for our
purposes, is given by the coset $SU(2,1)/U(2)$ whose natural metric is dual to
the standard Fubini-Study metric, 
$$ ds^2 =\fracmm{dr^2}{(1-r^2)^2}+\fracmm{r^2}{(1-r^2)^2}\s^2_2
+\fracmm{r^2}{(1-r^2)}(\s^2_1+\s^2_3)~.\eqno(2)$$
Note that the coefficient at $\s^2_2$ decays faster than the coefficients at  
$\s^2_1$ and  $\s^2_3$. The conformal structure associated with the metric (2) 
inside the unit ball in ${\bf C}^2$  survives  on the two-dimensional subspace
of the boundary of ${\bf C}^2$ that is annihilated by $\s_2$.

The metric (2) can be identified with the Zamolodchikov metric of certain 2d,
N=2 SCFT \cite{zam}. It is not difficult to identify the corresponding SCFT 
since the coset $SU(2,1)/U(2)$ appears in the Kazama-Suzuki coset construction
 list (see ref.~\cite{cftbook} for a review).  In fact, there is even 2d, N=4
supersymmetry in the 2d SCFT associated with this coset \cite{gk}.

\section{Weyl self-duality and RG flow}

The (Weyl) ASD equations can be put into an equivalent form of the first-order
 system of Ordinary Differential Equations (ODE) that are going to be
identified with the RG flow equations in 2d QFT. We are thus led to a study of
the $SU(2)$-invariant deformations of the metric (2) subject to the 
constraints (1). This well-defined mathematical problem was already addressed 
by Tod \cite{tod} and Nitchin \cite{hit}. A generic $SU(2)$ invariant metric 
in the Bianchi IX formalism reads
$$ ds^2=w_1w_2w_3dt^2+\fracmm{w_2w_3}{w_1}\s^2_1+\fracmm{w_3w_1}{w_2}\s^2_2+  
\fracmm{w_1w_2}{w_3}\s^2_3~.\eqno(3)$$
Being applied to eq.~(3), the Weyl ASD conditions of eq.~(1) give rise to 
the ODE system \cite{tod}
$$ \dt{A}_1=-A_2A_3 +A_1(A_2+A_3)~, \quad {\rm and~cyclic~permutations}~,
\eqno(4)$$
where the dot means differentiation with respect to $t$, while $A_i$, 
$i=1,2,3$, are defined from the auxiliary ODE system
$$ \dt{w}_1=-w_2w_3 +w_1(A_2+A_3)~, \quad {\rm and~cyclic~permutations}~.
\eqno(5)$$

The metric (2) corresponds to the case when all $A_i$ vanish. The 
Einstein condition in eq.~(1) can be easlily satisfied by conformal rescaling 
of the (Weyl) ASD metric (see below). Having solved eq.~(4), its
solution can be substituted into eq.~(5). To solve the last equations, it is
convenient to change variables as \cite{tod}
$$ w_1=\fracmm{\O_1\dt{x}}{\sqrt{x(1-x)}}~,\quad
 w_2=\fracmm{\O_2\dt{x}}{\sqrt{x^2(1-x)}}~,\quad
 w_3=\fracmm{\O_3\dt{x}}{\sqrt{x(1-x)^2}}~~,\eqno(6)$$
where $\O_i$ are constrained by the algebaric relation
$$\O_2^2+\O_3^2-\O^2_1=\fracm{1}{4}~~.\eqno(7) $$
Equation (5) then takes the form \cite{tod,hit}
$$ \O_1'=-\fracmm{\O_2\O_3}{x(1-x)}~,\quad
 \O_2'=-\fracmm{\O_3\O_1}{x}~,\quad  \O_3'=-\fracmm{\O_1\O_2}{1-x}~,\eqno(8)
$$
where the prime denotes differentiation with respect to $x$. The constraint 
(7) is preserved under eq.~(8), so that the transformation (6) is consistent. 
In terms of the new variables $(x,\O_i)$, the Einstein condition of eq.~(1) on 
the metric in terms of the new variables,
$$ ds^2=e^{2u}\left[ \fracmm{dx^2}{x(1-x)}+\fracmm{\s^2_1}{\O^2_1}
+\fracmm{(1-x)\s^2_2}{\O^2_2}+\fracmm{x\s^2_3}{\O^2_3}\right]~,\eqno(9)$$
amounts to the algebaric relation 
$$ 96\k^2e^{2u}=\fracmm{8x\O^2_1\O^2_2\O^2_3+2\O_1\O_2\O_3(x(\O_1^2+\O_2^2)-
(1-4\O^2_3)(\O^2_2-(1-x)\O ^2_1))}{(x\O_1\O_2+2\O_3(\O_2^2-(1-x)\O^2_1))^2}~.
\eqno(10)$$

The ODE system (4), $\dt{A}_i=C_i^{jk}A_jA_k$, can be naturally interpreted as
the RG flow equations in the dual 2d QFT originating from the 2d SCFT in 
its UV-fixed point. The K\"ahler nature of this 2d QFT implies that it should 
be N=2 supersymmetric. The universal coefficients $C_i^{jk}$ can be identified
 with the (normalized) OPE coefficients of the 2d, N=2 SCFT.

The exact solutions to the ODE system (4) are known to be dictated by
 the particular Painlev\'e VI equation, 
whose parameters are all fixed by the quaternionic-K\"ahler property of the 
metric \cite{tod,hit},
$$\eqalign{
y''~=~&\fracmm{1}{2}\left( \fracmm{1}{y} 
+\fracmm{1}{y-1}+\fracmm{1}{y-x}\right)
(y')^2- \left( \fracmm{1}{x} +\fracmm{1}{x-1}+\fracmm{1}{y-x}\right)y'\cr
&~  +\fracmm{y(y-1)(y-x)}{x^2(x-1)^2}\left[
\fracmm{1}{8} -\fracmm{x}{8y^2}+\fracmm{x-1}{8(y-1)^2}
+\fracmm{3x(x-1)}{8(y-x)^2}\right] ~,\cr}\eqno(11)$$
where $y=y(x)$, and the primes denote differentiation with respect to $x$.

The equivalence between eqs.~(4) and (11) is established via eq.~(8) and
 the relations \cite{hit}
$$\eqalign{
\O^2_1~=~&\fracmm{(y-x)^2y(y-1)}{x(1-x)}\left(v-\fracmm{1}{2(y-1)}\right)\left(
v-\fracmm{1}{2y}\right)~,\cr
\O^2_2~=~&\fracmm{(y-x)y^2(y-1)}{x}\left(v-\fracmm{1}{2(y-x)}\right)\left(
v-\fracmm{1}{2(y-1)}\right)~,\cr
\O^2_3~=~&\fracmm{(y-x)y(y-1)^2}{(1-x)}\left(v-\fracmm{1}{2y}\right)\left(
v-\fracmm{1}{2(y-x)}\right)~,\cr}\eqno(12)$$
where the auxiliary variable $v$ is defined by the auxiliary equation
$$ y'=\fracmm{y(y-1)(y-x)}{x(x-1)}\left(2v-\fracmm{1}{2y}-\fracmm{1}{2(y-1)}
+\fracmm{1}{2(y-x)}\right)~.\eqno(13)$$

An exact solution to eq.~(11), leading to a {\it complete} (regular) metric, 
is known to be unique, while it can be expressed in terms of the
 standard  theta-functions $\vq_{\a}(z|\t)$, $\a=1,2,3,4$. In order to write 
down the solution $y(x)$ explicitly, the theta-function arguments should be 
related by $z=\fracm{1}{2}(\t -k)$, where $k$ is considered to be an arbitrary
 (real and positive) parameter. Their relation to $x$ is defined by
$x=\vq^4_3(0)/\vq^4_4(0)$, where the value of $z$ is explicitly indicated, 
as usual. The relevant exact solution to the Painlev\'e VI equation reads 
\cite{hit}
$$\eqalign{
y(x)~=~& \fracmm{\vq_1'''(0)}{3\p^2\vq^4_4(0)\vq_1'(0)}+\fracmm{1}{3}\left[
1+\fracmm{\vq_3^4(0)}{\vq^4_4(0)}\right] \cr
 &~ +\fracmm{\vq_1'''(z)\vq_1(z)-2\vq_1''(z)\vq_1'(z)+
2\p i(\vq_1''(z)\vq_1(z)-\vq_1'{}^2(z))}{2\p^2\vq_4^4(0)\vq_1(z)(\vq_1'(z)+
\p i\vq_1(z))}~.\cr}\eqno(14)$$ 
 
The parameter  $k>0$ describes the monodromy of the solution (14) around its 
essential singularities (branch points) $x=0,1,\infty$. This (non-abelian) 
monodromy is generated by the matrices (with the eigenvalues $\pm i$)
$$ M_1=\left( \begin{array}{cc} 0 & i \\ i & 0\end{array} \right)~,\quad
  M_2=\left( \begin{array}{cc} 0 & i^{1-k} \\ 
i^{1+k} & 0\end{array} \right)~,\quad
M_3=\left( \begin{array}{cc} 0 & i^{-k} \\ 
-i^{k} & 0\end{array} \right)~.\eqno(15)$$

The function (14) is meromorphic outside $x=0,1,\infty$, with the simple poles
at $\bar{x}_1,\bar{x}_2,\ldots$, where $\bar{x}_n\in (x_n,x_{n+1})$ and 
$x_n=x(ik/(2n-1))$ for each positive integer $n$. Accordingly, the metric is 
well-defined (complete) for $x\in (\bar{x}_n,x_{n+1}]$, i.e. in the unit ball 
with the origin at $x=x_{n+1}$ and the boundary at $x=\bar{x}_n$ \cite{hit}.
Near the boundary the metric (9) has the asymptotical behaviour 
$$\eqalign{
 ds^2~=~&\fracmm{dx^2}{(1-x)^2} + \fracmm{4}{(1-x)\cosh^2(\p k/2)}\s^2_1 \cr
~& +\fracmm{16}{(1-x)^2\sinh^2(\p k/2)\cosh^2(\p k/2)}\s^2_2\cr
&~ + \fracmm{4}{(1-x)\sinh^2(\p k/2)}\s^2_3+ ~~{\rm regular~terms}~.\cr}
\eqno(16)$$
It is clear from eq.~(16) that the coefficient at $\s^2_2$ vanishes faster than
the others, like in eq.~(2), so that there is the natural conformal structure,
$$ \sinh^2(\p k/2)\s^2_1 +\cosh^2(\p k/2)\s^2_3~, \eqno(17)$$
on the two-dimensional boundary annihilated by $\s_2$. The only relevant 
parameter $\tanh^2(\p k/2)$ in eq.~(17) represents the central charge of 
the 2d CFT on the boundary. In the interior of the ball we have the spectral 
flow with the mononotically decreasing `effective' central charge  (called 
$c$-function), in accordance with the c-theorem \cite{zam}. The RG evolution 
ends at another (IR) fixed point where the solution (14) has a removable pole. 

\section{Conclusion}

Local 4d, N=2 supersymmetry appears to be the sole source of the constraints 
(1) on the effective metric. The regular $SU(2)$-invariant metric solutions 
are also unique, being parametrized by the CFT central charge describing the 
monodromy of the `master' solution to the underlying Painlev\'e equation. Our 
geometrical description of the RG flow by eq.~(1) is manifestly covariant with
respect to arbitrary reparametrizations of the 2d QFT coupling constants. In
our explicit example of the holographic correspondence, the RG flow in a 2d, 
N=2 QFT is described by the ODE system (4) whose coefficients are the 
universal (normalized) OPE coeffients of the underlying CFT at the UV-fixed 
point of the QFT. Unlike the 2d, N=2 supersymmetric RG flow solutions found by
 Cecotti and Vafa \cite{cv}, our RG flow has an IR fixed point and, therefore,
it  can be interpreted as a domain-wall solution.

{\bf Acknowledgements.} I would like to thank the Organizers of the Fradkin
Memorial Conference for the nice meeting and the opportunity to contribute a 
talk. I am grateful to Dmitry Alekseevsky, Luis Alvarez-Gaum\'e, Hermann 
Nicolai, Paul Tod and Galliano Valent for useful discussions. I also thank the
 Laboratoire de Physique Theorique et Hautes Energies in Paris VI, the 
Max-Planck-Institut f\"ur Gravitationsphysik in Golm, and the 
Max-Planck-Institut f\"ur Mathematik in Bonn, for kind hospitality extended to
 me during a preparation of this paper.

\end{document}